\documentclass[11pt,a4paper]{article}
\pdfoutput=1
\usepackage{jheppub}
\usepackage{graphicx,color,natbib}
\usepackage{amsmath,amssymb,amsfonts}
\usepackage{caption}
\usepackage{subcaption}
\usepackage{float}
\newcommand{\bse}{\begin{subequations}}
\newcommand{\ese}{\end{subequations}}
\newcommand{\be}{\begin{equation}}
\newcommand{\ee}{\end{equation}}
\newcommand{\bea}{\begin{eqnarray}}
\newcommand{\eea}{\end{eqnarray}}
\newcommand{\ba}{\begin{array}}
\newcommand{\ea}{\end{array}}

\def\h{\frac{1}{2}}

\def\FF{{\mathcal{F}}}
\def\HH{{\mathcal{H}}}
\def\B{{\mathcal{B}}}
\def\ZZ{{\mathcal{Z}}}

\input amssym.def
\input amssym.tex

\title{Meson Life Time in the Anisotropic Quark-Gluon Plasma}

\author[a,b]{Mohammad Ali-Akbari}
\author[b]{Davood Allahbakhshi}
\affiliation[a]{Department of Physics, Shahid Beheshti University, G.C., Evin, Tehran 19839, Iran.}
\affiliation[b]{School of Particles and Accelerators, Institute for Research in Fundamental Sciences (IPM),
P.O.Box 19395-5531, Tehran, Iran}
\emailAdd{aliakbari@theory.ipm.ac.ir}
\emailAdd{allahbakhshi@ipm.ir}

\abstract{
In the hot (an)isotropic plasma the meson life time $\tau$ is defined as a time scale after which the meson dissociates. According to the gauge/gravity duality, this time can be identified with the inverse of the imaginary part of the frequency of the quasinormal modes, $\omega_I$, in the (an)isotropic black hole background. In the high temperature limit, we numerically show that at fixed temperature(entropy density) the life time of the mesons decreases(increases) as the anisotropy parameter raises. For general case, at fixed temperature we introduce a polynomial function for $\omega_I$ and observe that the meson life time decreases. Moreover, we realize that $(s/T^3)^6$, where $s$ and $T$ are entropy density and temperature of the plasma respectively, can be expressed as a function of anisotropy parameter over temperature. Interestingly, this function is a \emph{Pad\'{e} approximant}. }

\begin{document}

\maketitle

\tableofcontents

\section{Introduction}
A new phase of quantum chromodynamics, quark-gluon plasma (QGP), is produced at relativistic heavy ion collider (RHIC) or these days at large hardon collider (LHC) by colliding two heavy nuclei such as gold (Au) or lead (Pb), relativistically. Experimental observations imply that the plasma is strongly coupled \cite{Shuryak:2003xe} and hence the perturbative calculation is not trustworthy. Therefore non-perturbative methods such as gauge/gravity duality may be applied to explain various properties of the plasma.

The gauge/gravity duality claims that for certain strongly coupled gauge theories the dynamics of the quantum fields can be described by the dynamics of the classical fields living in a higher dimensional space-time \cite{CasalderreySolana:2011us}. In particular, ${\cal{N}}=4$ super Yang-Mills theory (SYM) in the limit of large colors $N$ and large but finite t'Hooft coupling $\lambda$, which is expected to behave in a similar way with the strongly coupled QGP, is dual to type IIb supergravity on $AdS_5\times S^5$ background \cite{Maldacena}. Similarly a thermal SYM theory corresponds to the supergravity in an AdS-Shwarzschild background where the temperature of the SYM theory is identified with the Hawking temperature of AdS black hole \cite{Witten:1998qj}. Moreover Mateos and Trancanelli have introduced an interesting generalization of this duality to the thermal and spatially anisotropic SYM theory \cite{Azeyanagi:2009pr, Mateos:2011tv}. 

In order to add matter (quark) in the fundamental representation of the corresponding gauge group, one needs to introduce a D-brane into the background in the probe limit \cite{Karch:2002sh}. The probe limit means that D-brane does not back-react the geometry. Then the asymptotic shape of the brane gives the mass and condensation of the matter field. In addition, the shape of the brane can be classified into two types, one is the Mikowski embedding (ME) and the other is black hole embedding (BE). While the ME does not see the horizon, the BE crosses it. Various aspects of these embeddings have been studied in the literature, for instance see \cite{Mateos:2007vn}.

The results reported in \cite{Hoyos:2006gb} show that the mesons living in the QGP can be described by quasinormal modes. They are considered as certain small fluctuations around the BE with a complex frequency. Therefore, they are unstable modes where the imaginary part of their frequencies is identified with the inverse of the meson life time. The question we would like to answer in this paper is how the anisotropy affects the mass of the meson and its life time.

\section{Quasinormal Modes}


The background we are interested in is an anisotropic solution of
the IIb supergravity equations of motion. This
solution in the string frame is given by \cite{Mateos:2011tv} %
\begin{eqnarray} \label{one}%
 ds^2&=&-\FF\B u^{-2}dt^2+u^{-2}(dx^2+dy^2)+\HH u^{-2}dz^2\nonumber \\
 &+&\FF^{-1}u^{-2}du^2+e^{\h\phi} d\Omega_5^2, \nonumber \\
 d\Omega_5^2&=&d\theta^2+\sin^2\theta d\Omega_3^2+\cos^2\theta d\varphi^2, \nonumber \\
 \chi&=&az,\ \ \phi=\phi(u),
\end{eqnarray} %
where $a$ is a constant. $\chi$ and $\phi$ are axion and dilaton fields,
respectively. $\HH$, $\FF$ and $\B$ depend only on the radial direction, $u$.
In terms of the dilaton field, they are %
\bse\label{three}\begin{align} %
\HH&=e^{-\phi},\\
 \label{ff} \FF&=  \frac{e^{-\frac{1}{2}\phi}\left[ a^2 e^{\frac{7}{2}\phi}(4u+u^2\phi')+16\phi'\right]}{4(\phi'+u\phi'')}\,,
 \\
 \frac{\B'}{\B}&=\frac{1}{24+10
 u\phi'}\left(24\phi'-9u\phi'^2+20u\phi''\right)\,,
\end{align}\ese %

In order to find the solution one needs to solve the equation of motion for dilaton field. Then the above equations for metric components and suitable boundary conditions will specify the solution. For more detail see \cite{Mateos:2011tv}. Note also that the solution also contains a self dual five-form field.

The function $\FF(u)$ in the temporal and radial components of the metric is the blackening factor. Therefore the horizon is located at $u=u_h$ where $\FF(u_h)=0$ and the
Hawking temperature is given by $T=-\frac{1}{4\pi}\FF'(u_h)\sqrt{\B(u_h)}$.
The boundary lies at $u=0$ and the
metric approaches $AdS_5\times S^5$
asymptotically. 
The coordinates of the spacetime where the gauge theory lives are $(t,x,y,z)$ where there is a $U(1)$ symmetry in the $xy$-plane. We call $x$ and $y$ the transverse directions and the
longitudinal direction is $z$. An anisotropy is clearly seen between
the transverse and longitudinal directions. The entropy density per unit volume in the $xyz$-directions is given by %
\be %
 s =\frac{\pi^2}{2}N^2\frac{e^{-\frac{5}{4}\phi(u_h)}}{\pi^3u_h^3}.
\ee %

In order to add the fundamental matter to the $SU(N)$ gauge theory we have to introduce a D7-brane into the anisotropic background in the probe limit. The probe limit means that the D7-brane does not modify the geometry. Flavour D7-branes in this background have been studied previously, for example see \cite{Patino:2012py}. In fact the open strings stretched between probe D7-brane and the D3-D7 system leading to the geometry \eqref{one} give rise to the matter in the fundamental representation of the gauge group. The dynamics of the open strings is described by the DBI action
\be %
S_{DBI}=-\tau_7\int d^8\xi\  e^{-\phi} \sqrt{\det(G_{ab}+2\pi\alpha'F_{ab})}.
\ee %
The D7-brane tension is $\tau_7$ where $\tau_7^{-1}=(2\pi)^7 l_s^8g_s$  and $G_{ab}=g_{MN}\partial_a X^M\partial_b X^N$ where in the large $N$ and t' Hooft coupling limits the D3-D7 system is replaced with $g_{MN}$ given by \eqref{one}. The D7-brane is extended along $t,x,y,z,u$ and wrapped around $S^3\subset S^5$. Although the four-form and the axion fields are non-zero in the background, in such an embedding the Chern-Simon action has no contribution to the action. The
shape of the brane is given by the transverse directions $\theta$ and $\varphi$ where we choose $\varphi$ to be zero. Since we do not like to study the effect of the gauge field living on the brane, we also set $A_a$ to be zero. Because of the translational symmetry of the metric components in $xyz$  directions and the rotational symmetry in $\Omega_3$ directions, we consider that $\theta$ depends on the radial direction and time as it is shown in \eqref{theta}. Therefore, the Lagrangian reduces to 
\bea\label{lagrangian} %
{\cal{L}}&&=e^{-\phi(u)}\frac{\cos^3\theta(u,t)}{u^5\sqrt{\FF}} \\
&&\times \sqrt{\ZZ^3\HH[\B\FF(1+u^2\FF \ZZ \theta'(u,t)^2)-u^2\ZZ \dot{\theta}(u,t)^2]}.\nonumber
\eea%

The physical parameters we are interested in can be found from the asymptotic solution to $\theta(u)$ equation of motion, 
$\theta_c(u)=\theta_0 u+\theta_2 u^3+ \dots$ \cite{Hoyos:2011us}, 
where $m=\frac{\theta_0}{2\pi\alpha'}$ is the mass of the fundamental matter and $c=\theta_2-\frac{1}{6}\theta_0^3$ corresponds to condensation that is proportional to $\langle \bar{\psi}\psi \rangle$. 

It is well known that the small fluctuations about the shape (the equilibrium configuration) of the probe branes represent the low spin mesons \cite{Hoyos:2006gb}. They are classified into two types according to their frequencies. In the MEs the normal modes, which are the fluctuations with discrete real frequencies, only exist. However, in the case of the BH embeddings, the fluctuations fall into the black hole and the corresponding frequencies, the so-called quasinormal modes, are complex. Applying the AdS/CFT corresponding, the meson will be dissociated in the QGP after the life time, which is given by the inverse of the imaginary part of the frequency \textit{i.e.} $\tau\propto\omega_I^{-1}$ \cite{Hoyos:2006gb}. In order to find the meson life time $\tau$, let us start with the following ansatz
\be\label{theta} %
 \theta(u,t)=\theta_c(u)+\epsilon\;e ^{i \omega t}\zeta(u), 
\ee %
where $\theta_c(u)$ is a time-independent solution of the equation of motion for $\theta(u,t)$ resulting from \eqref{lagrangian}. Substituting the above ansatz into the equation of motion for $\theta(u,t)$ and expanding it up to the first order in $\epsilon$, one finds a \textit{nonlinear} equation for $\theta_c(u)$ and a \textit{linearised} equation for $\zeta(u)$. The suitable boundary conditions to solve the nonlinear equation are $\theta_h=\theta_c(u_h)$ and $\theta'_c(u_h)$. The latter is fixed in terms of $\theta_h$ by using the equation of motion for $\theta_c(u)$. Therefore, we have a one parameter family of solutions for the background profile of the brane $\theta_c(u)$.

In order to find the quasinormal modes, one needs to solve the linear equation of motion for the $\zeta(u)$ by applying the following boundary conditions: modes which are ingoing at the horizon and have zero source term at the boundary. The analytic solutions to the near horizon equation for $\zeta(u)$ are %
\be\label{asymptotic 1} %
 \zeta(u)\approx e^{\pm i \frac{\omega}{T} Log(1-u/u_h)},
\ee %
where the +(-) sign corresponds to the ingoing(outgoing) modes. On the other hand the near boundary equation can be analytically solved and the solution is  
\be\label{asymptotic 2} %
\zeta(u)=\zeta_1\; u+\zeta_3 \;u^3 + ... \ \ .
\ee %
To find the quasinormal modes we have to force the source term, $\zeta_1$, to equal zero or equivalently $\zeta'(u)|_{u=0}=0$. 
Considering the field redefinition 
\be %
\zeta(u)= e^{+ i \frac{\omega}{T} Log(1-u/u_h)} \; \psi(u),
\ee %
one can see that $\psi(u)$ has the regular expansion 
\be %
\psi(u)=\psi_0 + \psi_1 (u - u_h)+\psi_2 (u - u_h)^2+...\ ,
\ee %
near the horizon. Since the equation for $\psi$ is linear, $\psi_0$ can be set to 1 and the other coefficients will be determined from the equation of motion for $\psi(u)$. Interpolating between two asymptotic solutions \eqref{asymptotic 1} and \eqref{asymptotic 2} is possible only by a set of discrete complex values of $\omega$ which can be found by some standard methods such as shooting method. We would like to emphasize that the meson in its ground state, corresponding to the first quasinormal mode, is considered in this paper.

\subsection{High Temperature Limit}
Fortunately in the high temperature limit, $T\gg a$, the anisotropic
solution has been analytically introduced in \cite{Mateos:2011tv}. In this limit, up to leading order in
$a$, the functions $\FF$, $\B$ and the dilaton field are given by %
\bse\begin{align}%
 \FF&=1-\frac{u^4}{u_h^4}+a^2\hat{\FF}_2(u)+..., \\
 \B&=1-\frac{a^2u_h^2}{24}\left(\frac{10u^2}{u_h^2+u^2}+\log(1+\frac{u^2}{u_h^2})\right)+...,\\
 \phi&=-\frac{a^2u_h^2}{4}\log(1+\frac{u^2}{u_h^2})+...,
\end{align}\ese
where %
\begin{eqnarray}%
 \hat{\FF}_2(u)&=&\frac{1}{24u_h^2}\bigg(8u^2(u_h^2-u^2)-10u^4\log2\nonumber \\
 &+&(3u_h^4+7u^4)\log(1+\frac{u^2}{u_h^2})\bigg),
\end{eqnarray}
The temperature and the entropy density of the solution in terms of the anisotropy parameter are %
\bse\begin{align} %
 T&=\frac{1}{\pi u_h}+\frac{(5\log2-2)u_h}{48\pi^2}a^2 + O(a^4), \\
 \label{entropy density} s&=\h N^2 \pi^2 T^3+\frac{N^2 T}{16}a^2 + O(a^4).
\end{align}\ese%
On the other hand at low temperature limit, \textit{i.e.} $a\gg T$, the entropy density is
\be \label{low entropy}%
s=c_{ent}N^2 \ a^{1/3} \ T^{8/3} + ...,
\ee %
where $c_{ent}\approx 3.2$ \cite{Mateos:2011tv}.

\begin{figure}
\centering
	\begin{subfigure}[t]{.45\textwidth}
		\includegraphics[width=\linewidth]{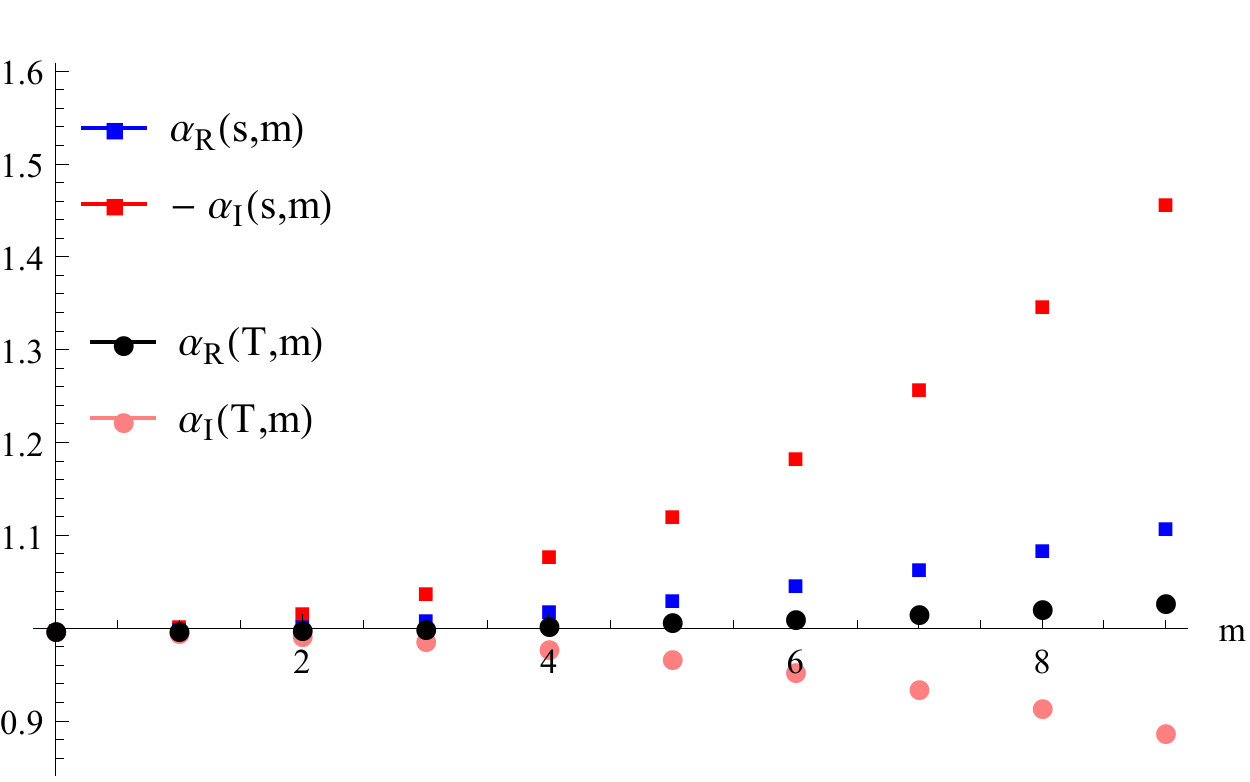}
		\caption{}
		\label{coefficient}
	\end{subfigure}
	\hspace{.5cm}
	\begin{subfigure}[t]{.45\textwidth}
		\includegraphics[width=\linewidth]{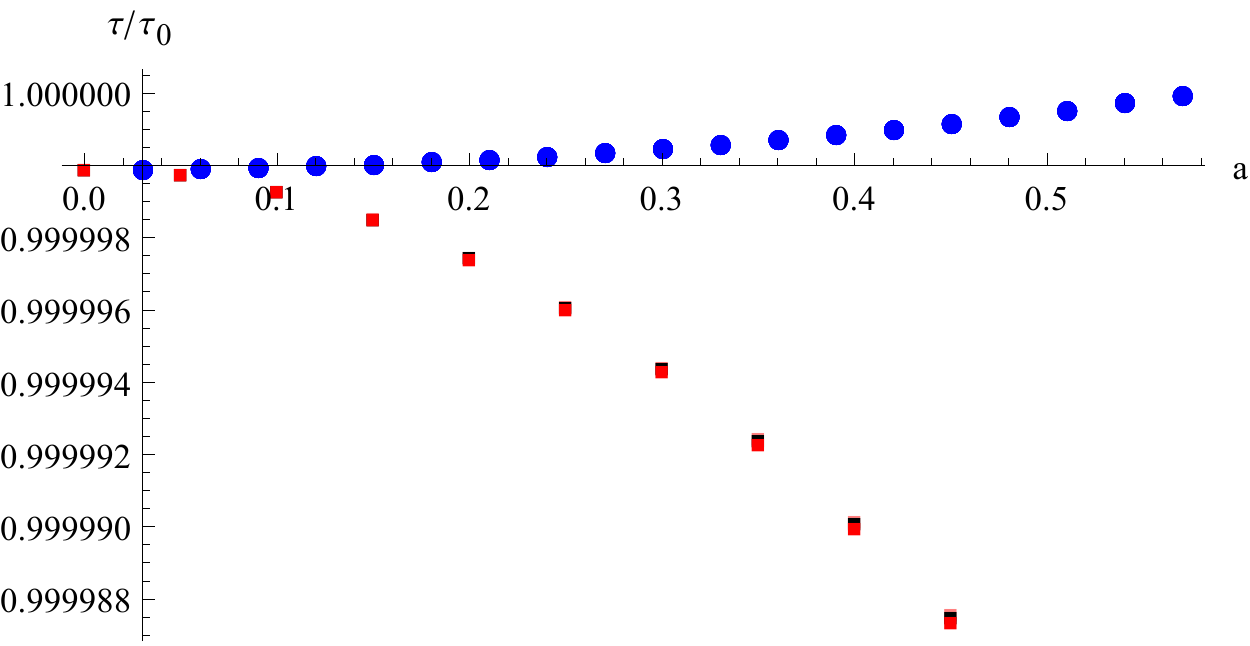}
		\caption{}
		\label{decay-time-scale}
	\end{subfigure}
\caption{Plot (a): At fixed temperature:  $\alpha_{R}$ and $\alpha_{I}$, normalized to 1 at $m=0$, have been plotted in terms of $m$. At fixed entropy density: $\alpha_{R}$ and $-\alpha_{I}$, normalized to 1 at $m=0$, have been plotted in terms of $m$. In this plot $\alpha_R(s/N^2=500\pi^2,0)=0.00129867$, 							$\alpha_I(s/N^2=500\pi^2,0)= -0.000347639$, $\alpha_R(T=10,0)=0.00418394$, $\alpha_I(T=10,0)= 0.00195894 $. Plot (b): Meson life time, normalized to 1 at a=0, has been plotted in terms of $a$, at fixed entropy density (blue points) and at fixed temperature (red squares).}
\end{figure}

We would like to find the frequency of the quasinormal modes in the high temperature background. Setting the anisotropy parameter equal to zero, both the real and imaginary parts of the frequency increase linearly as one raises the temperature. These are consistent with the results reported in \cite{Hoyos:2006gb}. As it is expected from the metric components at high temperature limit, $\omega_{R,I}$ depends on anisotropy parameter as $a^2$ for any given value of the temperature \textit{i.e.}
\be\label{frequency} %
 \omega_{R,I}=\omega^0_{R,I}(T,m) + \alpha_{R,I}(T,m)\;a^2,
\ee %
where $\omega^0_{R,I}(T,m)$ are frequencies of the quasinormal modes for the isotropic case \textit{i.e.} $a=0$.
Figure \ref{coefficient} shows that at fixed temperature although $\alpha_R$ is almost constant with increasing the mass of the faundamental matter, $\alpha_I$ decreases (a few values for $\omega^0_{R,I}$ are given in table \ref{table1}). It is important to notice that in a region around $m=T$ we expect a first order phase transition between black hole and Mikowski embeddings \cite{Herzog:2006gh} and therefore our results are not reliable in this region. Moreover, we numerically observe that for any given value of the mass a raise in the anisotropy parameter will increase $\omega_I$. And, in turn, as it is clearly seen from figure \ref{decay-time-scale}, it means that the $\tau/\tau_0$ decreases. Note that $\tau_0$ is the value of meson life time at $a=0$ for each corresponding mass. As a result the mesons will melt sooner in the QGP. This somehow indicates that anisotropy parameter and temperature behave similarly and it is in agreement with results in \cite{Ali-Akbari:2013txa,Chakraborty:2012dt}. We observe that the decrease in $\tau/\tau_0$ is almost the same for different masses. 

In the case of fixed entropy density, the behaviour of the real and imaginary parts of the frequency is similar to \eqref{frequency} as 
\be %
\omega_{R,I}=\omega^0_{R,I}(\frac{s}{N^2},m) + \alpha_{R,I}(\frac{s}{N^2},m)\;a^2,
\ee %
where its coefficients have been shown in the figure \ref{coefficient}. Compared to the mass dependence of $\alpha_{R,I}(T,m)$ at fixed temperature case, a notable increase can be seen for $\alpha_{R,I}(s/N^2,m)$.
Opposite to that seen in the fixed temperature case, raising the anisotropy in the system will increase the value of the $\tau/\tau_0$. 


\subsection{General Case}
In this section we are going to compute the real and imaginary parts of the frequency for arbitrary values of the temperature and anisotropy parameter. Our numerical computations show that both $\omega_R$ and $\omega_I$ grow linearly with increasing the temperature over a limited range of mass $0 < m < T$ when $a=0$. For instance, at fixed temperature, in the zero mass case we find
\bea\label{salam}
 \omega_R &=& 6.86 T+\delta\omega_R, \\
 \omega_I &=& 5.54 T+\delta\omega_I \nonumber.
\eea
The two second terms in above equations, $\delta\omega_R$ and $\delta\omega_I$, are a consequence of the anisotropy parameter.
The function for the deviations seems complicated but for example in massless case $\delta \omega_R$ and $\delta \omega_I$ may be approximated by the following polynomials
\bea\label{ahmad} %
 \delta\omega_{R,I}&=&\;T\;\sum_{n=2}^5 c_{n(R,I)} (\frac{a}{T})^n , \\
 c_{2R(I)}&=& 0.051(0.020),\ c_{3R(I)}=-0.0075(-0.002),\nonumber\\ 
 c_{4R(I)}&=& 4(0.6)\times 10^{-4},\ c_{5R(I)}=-6.86(0.68)\times 10^{-6}.\nonumber 
\eea %
Here we would like to emphasize that these functions can be applied in the range of our numerics ( $ 0.5 < T < 15 $ and $ 0 < a < 30 $ provided that $ a<9T $ ). 

\begin{table}
\begin{center}
$(T=10) \equiv (s/N^2=500 \ \pi^2)$\\
\begin{tabular}{c||c||c}
\hline
mass & $\omega^0_R(T,m)=\omega^0_R(s/N^2,m)$ & $\omega^0_I(T,m)=\omega^0_I(s/N^2,m)$\\
\hline
 0  & 68.7739 & 55.1979\\
 1  &  68.7727 & 55.207\\
 2  & 68.7694 & 55.2347\\
 3  & 68.76 & 55.2884\\
 4  & 68.7538 & 55.3537\\
 5  & 68.7502 & 55.4349\\
\end{tabular}
\caption{\label{table1} Isotropic frequencies for fixed values of temperature and entropy density in high temperature regime.}
\end{center}
\end{table}

One can also calculate the quasinormal modes when the entropy density is kept fixed. However, it is not easy to find suitable functions for $\delta\omega_{R,I}(s,m)$ which fit our numerical results. Instead, our data turn out to be fit with the following function 
\be\label{entropy} %
 s(a,T)=\frac{\pi ^2 }{2}N^2  T^3 \bigg(\frac{1+\alpha_2  \left(\frac{a}{T}\right)^2+\alpha_4  
 \left(\frac{a}{T}\right)^4}{1+\beta_2 \left(\frac{a}{T}\right)^2}\bigg)^{1/6}.
\ee
Using the expansion of the entropy density in the high temperature limit \cite{Mateos:2011tv}, $\alpha_4$ and $\beta_2$ can be obtained in terms of $\alpha_2$ as 
\be %
 \alpha_4=\frac{48\pi^2\alpha_2-37}{64 \pi^4},\ \beta_2= \frac{4\pi ^2\alpha_2-3}{4 \pi^2}. 
\ee %
Notice that this function gives the correct expression for the entropy density of ${\cal{N}}=4$ super Yang-Milles theory ($a=0$).
The value of $\alpha_2$ can be found by using the best fit for the entropy density and is obtained as $\alpha_2 = 1/4$ with an error less than 0.1\% (see fig \ref{padefig}). Surprisingly, this value leads to $c_{ent}\approx 3.205$ which is in perfect agreement with \eqref{low entropy}. Now \eqref{entropy} clearly leads to \eqref{low entropy} in the low temperature limit. The above discussion may be generalized by considering higher order terms. As a result we suggest %
\bea %
s(a,T)=
\frac{\pi ^2}{2}N^2  T^3 \bigg(\frac{1+\sum_{k=1}^{n+1} \alpha_{2k} (\frac{a}{T})^{2k}}{1+\sum_{k=1}^n \beta_{2k} (\frac{a}{T})^{2k}}\bigg)^{1/6},
\eea
which is a $[(2n+2)/2n]_f(a/T)$ \emph{Pad\'{e} approximant} for $f=(s/T^3)^6$. In principle, all coefficients can be achieved in terms of $\alpha_2$ by utilizing the higher order expansion of entropy density in terms of $a$ \cite{Mateos:2011tv}.

At fixed entropy density we found that the effect of anisotropy on the frequencies is very small (less than 1\%). In principle from \eqref{entropy}, for a fixed value of the entropy density and given $a$, the temperature can be found. Inserting the resultant temperature into \eqref{ahmad}, we obtain $\delta\omega_R$ and $\delta\omega_I$. Although it is promising that we can achive the real and imaginary parts of the frequecy at fixed entropy density, unfortunately the effect of anisotropy on the frequencies (1\%) is less than the error of the polynomials \eqref{ahmad} (4\%) and therefore the error washes away the effect.

\section{Discussion}

Main aim in this paper is to understand the effect of the anisotropy on the life time of the mesons living in the plasma. As it was already mentioned, according to gauge/gravity duality, the life time and the mass of the meson are described by $\omega_I^{-1}$ and $\omega_R$, respectively. By recalling \eqref{salam}, one can calculate the following ratios
\be\label{life time} %
\frac{\tau}{\tau_0}=\frac{\omega_I(0,T)}{\omega_I(a,T)}\ , \ \ \ \frac{M_{\rm{meson}}}{M^0_{\rm{meson}}}=\frac{\omega_R(a,T)}{\omega_R(0,T)}\ .
\ee %

\begin{figure}
\centering
	\begin{subfigure}[t]{.45\textwidth}
		\includegraphics[width=\linewidth]{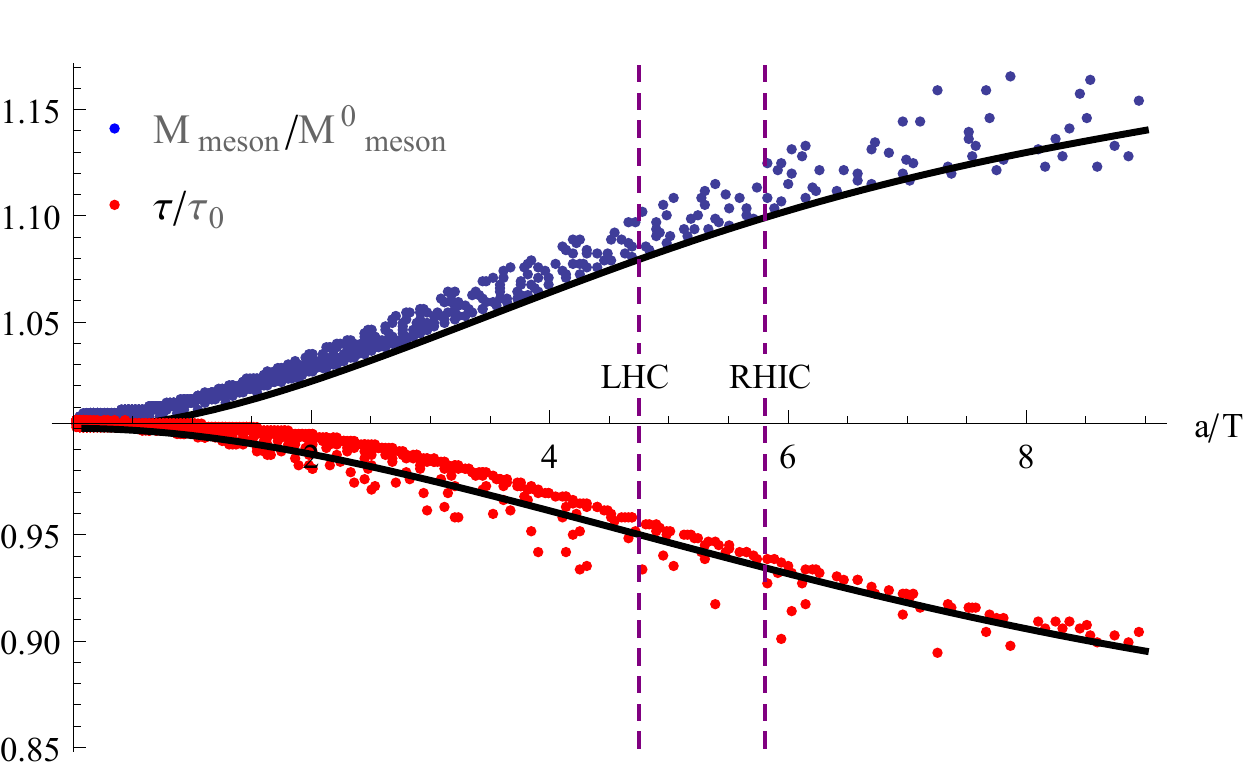}
		\caption{}
		\label{percent}
	\end{subfigure}
	\hspace{.5cm}
	\begin{subfigure}[t]{.45\textwidth}
		\includegraphics[width=\linewidth]{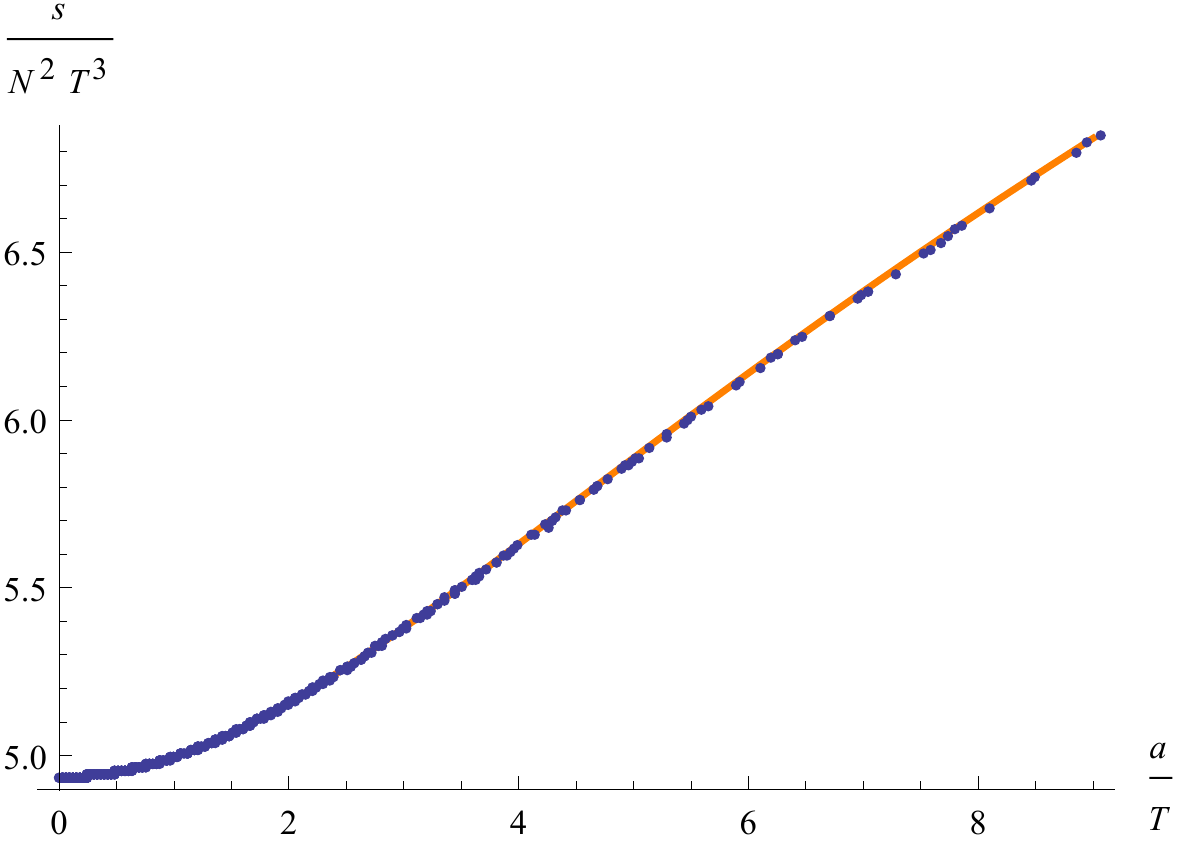}
		\caption{}
		\label{padefig}
	\end{subfigure}
\caption{Plot (a): The values of $M_{meson}/M^0_{meson}$ and $\tau / \tau_0$ has been plotted versus $a/T$. The dots are 
		  numerical data and solid curves are the fitted polynomials \eqref{ahmad}. Plot (b): The fitted function \eqref{entropy} (orange curve) and numerical data for $s/T^3$ (blue dots).}
\end{figure}

These ratios have been plotted in fig \ref{percent}. It was discussed in \cite{Giataganas:2012zy} that $a/T$ for RIHC(LHC) is $5.81(4.75)$. Using \eqref{salam} and \eqref{life time}, we then have
\bea %
 \ \rm{RHIC}\ (T=250 \rm{Mev})\left\{%
\begin{array}{ll} %
    \frac{M_{\rm{meson}}}{M^0_{\rm{meson}}}\approx 1.12,\\
    \frac{\tau}{\tau_0}\approx 0.90,
\end{array}%
\right.
\eea
and
\bea
\ \rm{LHC}\ (T=450 \rm{Mev})\left\{%
\begin{array}{ll} %
    \frac{M_{\rm{meson}}}{M^0_{\rm{meson}}}\approx 1.08,\\
    \frac{\tau}{\tau_0}\approx 0.92,
\end{array}%
\right.
\eea %
and therefore the mesons dissociate more in the presence of anisotropy. This conclusion is in agreement with the result reported in \cite{Chernicoff:2012bu}. In this paper it was shown that the screening length as a function of the anisotropy decreases indicating that the life time of the bound states become shorter in the anisotropic plasma. Furthermore, at RHIC(LHC) energies an increase in the mass of the mesons occurs which is about $12(8)\%$. Since the QGP produced in laboratory is intrinsically anisotropic, one can not measure the mass of the meson living in the QGP for $a=0$. But, interestingly, this mass can be eliminated from our results and we then have %
\be %
 \frac{({M_{\rm{meson}})_{\rm{RHIC}}}}{(M_{\rm{meson}})_{\rm{LHC}}}\approx 1.037 
\ee  
In other words the effect of anisotropy can experimentally be observed by comparing the mass of the meson at RHIC and LHC. In fact at LHC energies, the meson is lighter.


\textit{\textbf{Acknowledgement:}}
We would like to thank A. Davody and H. Ebrahim for fruitful discussions.

\end{document}